\documentclass[aps,nofootinbib,showpacs,preprint]{revtex4}
\usepackage{graphicx}
\usepackage{amsfonts}

\usepackage{color}

\begin{document}


\title{Twofold Hidden Conformal Symmetries \\ of the Kerr-Newman Black Hole}

\author{Chiang-Mei Chen} \email{cmchen@phy.ncu.edu.tw}
\affiliation{Department of Physics and Center for Mathematics and Theoretical Physics,
National Central University, Chungli 320, Taiwan}

\author{Ying-Ming Huang} \email{y.m.huang26@gmail.com}
\affiliation{Department of Physics, National Central University, Chungli 320, Taiwan}

\author{Jia-Rui Sun} \email{jrsun@phy.ncu.edu.tw}
\affiliation{Department of Physics, National Central University, Chungli 320, Taiwan}

\author{Ming-Fan Wu} \email{93222036@cc.ncu.edu.tw}
\affiliation{Department of Physics, National Central University, Chungli 320, Taiwan}

\author{Shou-Jyun Zou} \email{sgzou2000@gmail.com}
\affiliation{Department of Physics, National Central University, Chungli 320, Taiwan}

\date{\today}


\begin{abstract}
In this paper, we suggest that there are two different individual 2D
CFTs holographically dual to the Kerr-Newman black hole, coming from
the corresponding two possible limits --- the Kerr/CFT and
Reissner-Nordstr\"om/CFT correspondences, namely there exist the
Kerr-Newman/CFTs dualities. A probe scalar field at low frequencies
turns out can exhibit two different 2D conformal symmetries (named
by $J$- and $Q$-pictures, respectively) in its equation of motion
when the associated parameters are suitably specified. These twofold
dualities are supported by the matchings of entropies, absorption
cross sections and real time correlators computed from both the
gravity and the CFT sides. Our results lead to a fascinating
``microscopic no hair conjecture''
--- for each macroscopic hair parameter, in additional to the mass of a
black hole in the Einstein-Maxwell theory,
there should exist an associated holographic
CFT$_2$ description.
\end{abstract}


\maketitle
\tableofcontents

\section{Introduction}
The holographic principle~\cite{'tHooft:1993gx, Susskind:1994vu} and
its first realization in string theory, i.e. the AdS$_5$/CFT$_4$
correspondence~\cite{Maldacena:1997re,Gubser:1998bc,Witten:1998qj}
provided us new perspectives and methods in finding quantum theory
of gravity. Recent major advances made along this direction was
began by the searching of quantum gravity descriptions of the Kerr
black hole, namely the Kerr/CFT correspondence~\cite{Guica:2008mu},
together with its various generalizations~\cite{Dias:2009ex,
Matsuo:2009sj, Bredberg:2009pv, Amsel:2009pu, Hartman:2009nz,
Castro:2009jf, Cvetic:2009jn, Hartman:2008pb, Garousi:2009zx,
Chen:2009ht, Chen:2010bs, Hotta:2008xt, Lu:2008jk, Azeyanagi:2008kb,
Chow:2008dp, Azeyanagi:2008dk, Nakayama:2008kg, Isono:2008kx,
Peng:2009ty, Chen:2009xja, Loran:2009cr, Ghezelbash:2009gf,
Lu:2009gj, Amsel:2009ev, Compere:2009dp, Krishnan:2009tj,
Hotta:2009bm, Astefanesei:2009sh, Wen:2009qc, Azeyanagi:2009wf,
Wu:2009di, Peng:2009wx, Chen:2009cg, Chen:2010ni, Becker:2010jj,
Balasubramanian:2009bg, Castro:2010vi}. The Kerr/CFT correspondence
was first proposed for the near horizon extremal Kerr black hole
which contains an AdS$_2 \times S^1$ structure (a warped AdS$_3$
structure with $SL(2, R)_R \times U(1)_L$ isometry), then the left
hand central charge $c_L = 12 J$ of the dual 2D CFT was obtained by
following similar treatment in the AdS$_3$/CFT$_2$
duality~\cite{Brown:1986nw} to analyze the asymptotical symmetries
of the spacetime, where $J$ is the angular momentum of the Kerr
black hole. It was later shown that the Kerr/CFT duality can be
generalized to the near extremal Kerr black hole case by studying
the low frequency scattering process of external
fields~\cite{Bredberg:2009pv, Hartman:2009nz}. For both the extremal
and near extremal Kerr black holes, the near horizon AdS$_2 \times
S^1$ geometries play an essential role in obtaining the central
charges and conformal weights of the dual 2D CFTs. However, when the
black hole is non-extremal, there is no apparent near horizon AdS
structures, thus the usual AdS/CFT approaches do not work directly.
Remarkably, it was suggested that a generic non-extremal Kerr black
hole should still dual to a 2D CFT based on the fact that there
exists a hidden 2D conformal symmetry which can be probed by a
scalar field at low frequencies~\cite{Castro:2010fd}. For other
extensions on the hidden conformal symmetry, see~\cite{Chen:2010as,
Krishnan:2010pv, Wang:2010qv, Rasmussen:2010sa, Chen:2010xu,
Li:2010ch, Chen:2010zw, Krishnan:2010df, Rasmussen:2010xd,
Becker:2010dm, Chen:2010bh, Wang:2010ic, Chen:2010yu}.

Among the generalizations of the Kerr/CFT correspondence, an
interesting progress is the studying of the CFT description of the
Reissner-Nordstr\"om (RN) black hole, i.e. the RN/CFT
duality~\cite{Hartman:2008pb, Garousi:2009zx, Chen:2009ht,
Chen:2010bs}. Since the RN black hole is non-rotating and it only
contains an AdS$_2$ geometry in the near horizon (near) extremal
limit, to follow the method of the Kerr/CFT correspondence, one can
uplift the 4D RN black hole into a five dimensional spacetime, in
which a $U(1)$ symmetry in the 5D metric appears and then a (warped)
AdS$_3$ structure comes out in the (near) extremal near horizon
region~\cite{Hartman:2008pb, Garousi:2009zx, Chen:2010bs}. An
alternative way is to reduce the 4D RN black hole into two
dimensions and investigate the stress tensors and current of its 2D
effective theory by analyzing the asymptotical
symmetries~\cite{Chen:2009ht}. Like the Kerr black hole case, the
RN/CFT correspondence is found not only valid for the extremal and
near extremal cases, but also for the non-extremal RN black hole,
motivated by the work of probing hidden conformal symmetry.
Moreover, we showed that for the 4D RN black hole, like its 5D
uplifted counterpart, its holographic dual CFT should still be two
dimensional~\cite{Chen:2010as, Chen:2010yu}. A key observation made
in~\cite{Chen:2010yu} is that the $U(1)$ symmetry of the background
electromagnetic field can be probed by a charged scalar field, and
then the hidden 2D conformal symmetry of the 4D RN black hole can be
revealed. Consequently, a dual CFT$_2$ description is conjectured.
The result in~\cite{Chen:2010yu} indicates that the $U(1)$ symmetry of
the background gauge field plays an equivalent role with that of the
$U(1)$ symmetry coming from the rotation. This is actually part of
the motivations for our present paper on exploring the twofold
hidden conformal symmetries of the Kerr-Newman (KN) black hole.

With the Kerr/CFT and RN/CFT correspondences in hand, naturally the
next step is to investigate the CFT duals for the KN black hole.
There were some trials on this problem but the results are
incomplete~\cite{Hartman:2008pb, Wang:2010qv, Chen:2010bh}. The
reason is that in all the above attempts, only the angular momentum
($J$-) picture of the KN black hole was revealed, where the central
charges of the dual 2D CFT are $c_L = c_R = 12J$, which do not
depend on the black hole charge $Q$. Although this $J$-picture can
reproduce correct entropy of the KN black hole and other information
of the dual CFT, it is still unsatisfied. Since geometrically, the
KN black hole will return to the Kerr black hole when $Q = 0$ while
to the RN black hole when $J = 0$. However, if one takes $J = 0$ in
the $J$-picture, it will not recover the information of the
corresponding RN black hole. That is to say, a multiple CFTs dual to
the KN black hole is expected, more specifically, another
$Q$-picture (in which the central charges will depend on $Q$) should
exist for the KN black hole. This is another part of motivations for
our present work. In this paper we will show that, there are indeed
two different individual 2D CFTs dual to the generic non-extremal KN
black hole, based on the fact that there are twofold 2D conformal
symmetries in the low frequency wave equation of the external scalar
field when we suitably turn on/off the couplings. Though these 2D
conformal symmetries are not derived from the geometry of KN black
hole background except for near horizon (near) extremal limits.
Nevertheless, they should reveal the hidden conformal symmetries of
the KN black hole and could still be understood in a geometric
picture which will be discussed in the section of Conclusion. Our
result indicates that each of the dual CFT gives a complete
holographic description for the KN black hole, which can be
described as a KN/CFTs dualities. Our suggestion is further
supported by matching of the absorption cross sections and real time
correlators calculated in both the $J$- and the $Q$-pictures from
the gravity and the CFT sides. In addition, our work also leads to a
fascinating ``microscopic no hair conjecture''
--- for each macroscopic hair parameter, in additional to the mass of a
black hole in the Einstein-Maxwell theory (with
dimension $D\geq 3$), there should exist an associated holographic
CFT$_2$ description.

The outline of this paper is as follows. We first review some basic
properties of the KN black hole and study the scattering process of
a probe charged scalar field propagating in its background in
section II. In section III, we will reanalyze the angular momentum
$J$-picture of the KN black hole in order to compare it with the
$Q$-picture. This section include the probing of hidden conformal
symmetry, the calculation of absorption cross sections and real time
correlators. Then in section IV, we explore explicitly the charge
$Q$-picture of the KN black hole parallel to the $J$-picture. We
give the conclusion in section V, and Appendix A list the symmetry
and Casimir operators in AdS$_3$ spacetime.

\section{Charged Scalar Field in the Kerr-Newman Background}
The general or unique axisymmetric black hole solution of the
four-dimensional Einstein-Maxwell theory
\begin{equation}
I = \frac1{16\pi} \int d^4x \sqrt{-g} \left( R - F_{[2]}^2 \right).
\end{equation}
is the Kerr-Newman (KN) black hole which are characterized by three
macroscopic quantities (hairs): mass $M$, electric charge $Q$ and
angular momentum $J = a M$. In the Boyer-Lindquist coordinates, the
KN black hole is expressed as
\begin{eqnarray}
ds^2
&=& - \frac{\Delta - a^2 \sin^2\theta}{\Sigma} \left[ dt \!+\! \frac{(2 M r \!-\! Q^2) a \sin^2\theta}{\Delta - a^2 \sin^2\theta} d\phi \right]^2 + \Sigma \left( \frac{dr^2}{\Delta} \!+\! d\theta^2 \!+\! \frac{\Delta \sin^2\theta}{\Delta - a^2\sin^2\theta} d\phi^2 \right),
\nonumber\\
A_{[1]} &=& - \frac{Q r}{\Sigma} \left( dt - a \sin^2\theta d\phi \right),
\end{eqnarray}
where
\begin{equation}
\Sigma = r^2 + a^2 \cos^2\theta, \qquad \Delta = r^2 - 2 M r + a^2 + Q^2.
\end{equation}
The radii of black hole outer and inner horizons $r_\pm$, the
horizon angular velocity $\Omega_H$ and the chemical potential
$\Phi_H$ are
\begin{equation}
r_\pm = M \pm \sqrt{M^2 - a^2 - Q^2}, \qquad \Omega_H = \frac{a}{r_+^2 + a^2}, \qquad \Phi_H = \frac{Q r_+}{r_+^2 + a^2},
\end{equation}
and the corresponding thermodynamical quantities, i.e. the Hawking
temperature and the black hole entropy, are
\begin{equation}
T_H = \frac{\kappa}{2 \pi} = \frac{r_+ - r_-}{4\pi (r_+^2 + a^2)}, \qquad S_{BH} = \frac{A_+}{4} = \pi (r_+^2 + a^2),
\end{equation}
where $\kappa$ and $A_+$ are the surface gravity and area of the
outer horizon, respectively.

Consider a probe charged scalar field scattering in the the KN black
hole background, the admitted motion is described by the
Klein-Gordon (KG) equation
\begin{equation}
(\nabla_{\alpha} - i q A_{\alpha})(\nabla^{\alpha} - i q A^{\alpha})
\Phi = -\mu^2\Phi,
\end{equation}
where $\mu$ and $q$ are mass and electric charge parameters of the
scalar field respectively. After assuming the following mode
expansions
\begin{equation}
\Phi(t, r, \theta, \phi) = \mathrm{e}^{- i \omega t + i m \phi} R(r) S(\theta),
\end{equation}
the KG equation can be decoupled into the angular and radial
equations as
\begin{eqnarray}
\frac1{\sin\theta} \partial_\theta (\sin\theta \, \partial_\theta S) - \left[ a^2 (\omega^2 - \mu^2) \sin^2\theta + \frac{m^2}{\sin^2\theta} - \lambda \right] S &=& 0
\\
\partial_r (\Delta \partial_r R) + \left[ \frac{[ (r^2 + a^2) \omega - q Q r - m a ]^2}{\Delta} - \mu^2 (r^2 + a^2) + 2 m a \omega - \lambda \right] R &=& 0,
\end{eqnarray}
where $\lambda$ is the separation constant. Furthermore, for a
massless probe scalar field with $\mu = 0$, the radial equation can
be further expressed in the following desirable form
\begin{eqnarray}\label{radial}
\partial_r (\Delta \partial_r R) \!+\! \left[ \frac{\left[ (r_+^2 \!+\! a^2) \omega - a m - Q r_+ q \right]^2}{(r - r_+) (r_+ - r_-)} \!-\! \frac{\left[ (r_-^2 \!+\! a^2) \omega - a m - Q r_- q \right]^2}{(r - r_-) (r_+ - r_-)} \right] R &&
\nonumber\\
+ \left[ \omega^2 r^2 + 2 (\omega M - q Q) \omega r +  \omega^2 a^2 - \omega^2 Q^2 + (2 \omega M - q Q)^2 \right] R &=& \lambda R.
\end{eqnarray}
Apparently, the potential terms in the second line can be neglected
when we imposing the following conditions: (1) small frequency
$\omega M \ll 1$ (consequently $\omega a \ll 1$ and $\omega Q \ll
1$), (2) small probe charge $q Q \ll 1$ and (3) near region $\omega
r \ll 1$. Moreover, the condition $\omega^2 a^2 \ll 1$ simplifies
the solution of the angular equations to be the Legendre polynomials
and the separation constant should have the standard values $\lambda
= l (l + 1)$. Finally, the radial equation (\ref{radial}) reduces to
\begin{equation} \label{eqR}
\partial_r (\Delta \partial_r R) + \left[ \frac{\left[ (r_+^2 + a^2) \omega - a m - Q r_+ q \right]^2}{(r - r_+) (r_+ - r_-)} - \frac{\left[ (r_-^2 + a^2) \omega - a m - Q r_- q \right]^2}{(r - r_-) (r_+ - r_-)} \right] R = l (l + 1) R.
\end{equation}

Remarkably, there are twofold 2D conformal symmetries encoded in the
solution space of the above equation~(\ref{eqR}). As we will see,
the affiliated coordinates of the operators $\partial_r$ and
$\partial_t$ (originating the frequency $\omega$) compose the two
dimensional base of the apparent AdS$_3$ structure of the searching
hidden conformal symmetry. However, there are two possible
candidates for the fibration --- either the $\phi$-coordinate
associated with the mode parameter $m$ or the internal coordinate of
the $U(1)$ symmetry space connected with the electric charge $q$ of
the probe scalar field. It turns out that each choice of fibration
coordinate gives a different apparent AdS$_3$ and therefore,
individual CFT$_2$ description. For the choice of $\phi$ coordinate,
the corresponding field theory is an extension of the CFT dual to
the Kerr black hole --- with the same central charges determined by
the angular momentum but generalized temperatures. This description,
called the $J$-picture, has been well-investigated recently
in~\cite{Wang:2010qv, Chen:2010xu} and we are going to reproduce the
results, for comparison with another picture, in this paper. The
other choice of fibration, i.e. the internal $U(1)$ phase, similarly
gives a generalization of the RN/CFT correspondence in which the CFT
central charges are given by the charge of the black hole. The
realization of the second description, called $Q$-picture, is not so
obvious by using the technique developed in~\cite{Chen:2010yu} for the
dyonic RN black hole. All the details will be presented in the
coming sections.

\section{Angular Momentum ($J$-) Picture}

\subsection{Hidden Conformal Symmetry}
In order to probe the $J$-picture CFT$_2$ description dual to the KN black hole, one should consider a neutral scalar field by imposing $q = 0$ as a probe and the equation~(\ref{eqR}) becomes to
\begin{equation}
\left( \partial_r (\Delta \partial_r) - \frac{\left[ (r_+^2 + a^2) \partial_t + a \partial_\phi \right]^2}{(r - r_+) (r_+ - r_-)} + \frac{\left[ (r_-^2 + a^2) \partial_t + a \partial_\phi \right]^2}{(r - r_-) (r_+ - r_-)} \right) \Phi = l (l + 1) \Phi.
\end{equation}
By identifying the coordinate $\phi$ as $\chi$, then the left hand side of the above equation is just the Casimir operator~(\ref{Casimir}) acting on $\Phi$ with the following relations of parameters
\begin{equation}
T^J_L = \frac{r_+^2 + r_-^2 + 2 a^2}{4 \pi a (r_+ + r_-)}, \quad T^J_R = \frac{r_+ - r_-}{4 \pi a}, \qquad n^J_L = - \frac1{2 (r_+ + r_-)}, \quad n^J_R = 0.
\end{equation}
These results were derived recently in~\cite{Wang:2010qv, Chen:2010xu}. In this picture, the probe scalar field does not couple with the background gauge field, so it reveals essentially the angular momentum dominated section of the dual field theory, resembling the dual CFT of the Kerr black hole. The corresponding central charges are completely determined by the angular momentum
\begin{equation}
c^J_L = c^J_R = 12 J = 12 M a.
\end{equation}
As the first supporting evidence, one can straightforwardly check
that the CFT microscopic entropy from the Cardy formula indeed
reproduces the Bekenstein-Hawking area entropy of the KN black hole:
\begin{equation}
S^J_\mathrm{CFT} = \frac{\pi^2}3 \left( c^J_L T^J_L + c^J_R T^J_R \right) = \pi (r_+^2 + a^2) = S_\mathrm{BH}.
\end{equation}

\subsection{Scattering}
For the further evidences to support the $J$-picture in KN/CFTs correspondence, we study the scattering process of the neutral probe scalar field in the KN black hole background. We can compute, from the gravity side, the absorption cross section and the real time correlator which can be verified in agreement with the corresponding results for the operator dual to the scalar field in the 2D CFT.

The near region KG equation~(\ref{eqR}) can be solved technically
easier in terms of a new variable $z$ defined by
\begin{equation}\label{defZ}
z = \frac{r - r_+}{r - r_-},
\end{equation}
and the general solutions, after imposing $q = 0$, include ingoing and outgoing modes as
\begin{eqnarray}
R_J^\mathrm{(in)} &=& z^{- i \gamma_J} (1 - z)^{l + 1} \, F( a_J, b_J; c_J; z ),
\nonumber\\
R_J^\mathrm{(out)} &=& z^{i \gamma_J} (1 - z)^{l + 1} \, F( a_J^*, b_J^*; c_J^*; z ),
\end{eqnarray}
in terms of the hypergeometric function $F( a, b; c; z )$ labeled by the following coefficients
\begin{eqnarray}
\gamma_J &=& \frac{\omega (r_+^2 + a^2) - m a}{r_+ - r_-},
\nonumber\\
a_J &=& 1 + l - i \frac{\omega (r_+^2 + r_-^2 + 2 a^2) - 2 m a}{r_+ - r_-},
\nonumber\\
b_J &=& 1 + l - i \omega (r_+ + r_-),
\nonumber\\
c_J &=& 1 - i 2 \gamma_J.
\end{eqnarray}
There is an important relation among these coefficients: $c_J - a_J - b_J = - 2 l - 1$. Actually the ingoing mode, more precisely its asymptotically behavior at the match region $r \gg M$ (but still satisfying $r \ll 1/\omega$) already contained the necessary information for the absorption cross section and real time correlator. The asymptotic form can be read out by taking the limits $z \to 1$ and $1 - z \to r^{-1}$, and the result for the  ingoing mode is
\begin{equation}\label{asymRJ}
R_J^\mathrm{(in)}(r \gg M) \sim A_J \, r^l + B_J \, r^{- l - 1},
\end{equation}
where two coefficients are
\begin{equation}
A_J = \frac{\Gamma(c_J) \Gamma(2 l + 1)}{\Gamma(a_J) \Gamma(b_J)}, \qquad B_J = \frac{\Gamma(c_J) \Gamma(- 2 l - 1)}{\Gamma(c_J - a_J) \Gamma(c_J - b_J)}.
\end{equation}
One more additional information can be obtained is that the conformal weights of the operator dual to the scalar field should be
\begin{equation}
h^J_L = h^J_R = l + 1.
\end{equation}
Hence, the coefficients $a_J$ and $b_J$ can be expressed in terms of parameters $\omega^J_L$ and $\omega^J_R$
\begin{equation}
a_J = h^J_R - i \frac{\omega^J_R}{2 \pi T^J_R}, \qquad b_J = h^J_L - i \frac{\omega^J_L}{2 \pi T^J_L},
\end{equation}
with
\begin{eqnarray}
\omega^J_L &=& \frac{\omega (r_+^2 + r_-^2 + 2 a^2)}{2 a},
\nonumber\\
\omega^J_R &=& \frac{\omega (r_+^2 + r_-^2 + 2 a^2) - 2 m a}{2 a}.
\end{eqnarray}
The essential part of the absorption cross section can be read out directly from the coefficient $A_J$, namely
\begin{equation}
P^J_\mathrm{abs} \sim |A_J|^{-2} \sim \sinh( 2 \pi \gamma_J) \, \left| \Gamma(a_J) \right|^2 \, \left| \Gamma(b_J) \right|^2.
\end{equation}

In order to compare the absorption cross section with the two-point function of the operator dual to the probe scalar field, one needs to identify the conjugate charges, $\delta E^J_L$ and $\delta E^J_R$, defined by
\begin{equation}
\delta S^J_{CFT} = \frac{\delta E^J_L}{T^J_L} + \frac{\delta E^J_R}{T^J_R}.
\end{equation}
from the first law of black hole thermodynamics
\begin{equation} \label{1st}
\delta S_{BH} = \frac1{T_H} \delta M - \frac{\Phi_H}{T_H} \delta Q - \frac{\Omega_H}{T_H} \delta J.
\end{equation}
One can get the the conjugate charges via $\delta S^J_{CFT} = \delta S_{BH}$ and the solutions are
and the solution~\cite{Chen:2010xu}
\begin{eqnarray}
\delta E^J_L &=& \frac{(2M^2 - Q^2) M}{J} \delta M - \left( \frac{M^2 Q}{J} - \frac{Q^3}{2 J} \right) \delta Q,
\nonumber\\
\delta E^J_R &=& \frac{(2M^2 - Q^2) M}{J} \delta M - \frac{M^2 Q}{J} \delta Q - \delta J.
\end{eqnarray}
The identifications of parameters are $\delta M = \omega, \delta Q = q$ and $\delta J = m$. Since the probe scalar field is neutral, in such case, we  have
\begin{equation}
\omega^J_L = \delta E^J_L(\delta M = \omega, \delta J = m; \delta Q = 0), \qquad \omega^J_R = \delta E^J_R(\delta M = \omega, \delta J = m; \delta Q = 0).
\end{equation}

Therefore, one can straightforwardly verify the following relation for the imaginary part of the coefficient $c_J$,
\begin{equation}
2 \pi \gamma_J = \frac{\omega^J_L}{2 T^J_L} + \frac{\omega^J_R}{2 T^J_R}.
\end{equation}
Finally, the absorption cross section can be expressed as
\begin{equation}
P^J_\mathrm{abs} \sim (T^J_L)^{2 h^J_L - 1} (T^J_R)^{2 h^J_R - 1} \sinh\left( \frac{\omega^J_L}{2 T^J_L} + \frac{\omega^J_R}{2 T^J_R} \right) \left| \Gamma\left( h^J_L + i \frac{\omega^J_L}{2 \pi T^J_L} \right) \right|^2 \, \left| \Gamma\left( h^J_R + i \frac{\omega^J_R}{2 \pi T^J_R} \right) \right|^2,
\end{equation}
which is the finite temperature absorption cross section of an operator dual to the neutral probe scalar field in the $J$-picture 2D CFT corresponding to the KN black hole.

\subsection{Real-time Correlator}
One can further check the real-time correlator~\cite{Son:2002sd} in the $J$-picture. The asymptotic behavior of ingoing scalar field~(\ref{asymRJ}) indicates that two coefficients play different roles: $A_J$ as the source and $B_J$ as the response, and the two-point retarded correlator is simply~\cite{Chen:2010ni}
\begin{equation}
G^J_R \sim \frac{B_J}{A_J} = \frac{\Gamma(- 2 l - 1)}{\Gamma(2 l + 1)} \frac{\Gamma(a_J) \Gamma(b_J)}{\Gamma(c_J - a_J) \Gamma(c_J - b_J)},
\end{equation}
Following the identity $c_J - a_J - b_J = - 2 l - 1$, we can easily check that the retarded Green function is
\begin{equation}
G^J_R \sim
\frac{\Gamma\left( h^J_L - i \frac{\omega^J_L}{2 \pi T^J_L} \right) \Gamma\left( h^J_R - i \frac{\omega^J_R}{2 \pi T^J_R} \right)}{\Gamma\left( 1 - h^J_L - i \frac{\omega^J_L}{2 \pi T^J_L} \right) \Gamma\left( 1 - h^J_R - i \frac{\omega^J_R}{2 \pi T^J_R} \right)}.
\end{equation}
Using the relation $\Gamma(z) \Gamma(1-z) = \pi/\sin(\pi z)$ we have
\begin{eqnarray}\label{GreenRJ}
G^J_R &\sim& \sin\left(\pi h^J_L + i \frac{\omega^J_L}{2 T^J_L} \right) \sin\left(\pi h^J_R + i \frac{\omega^J_R}{2 T^J_R} \right)
\nonumber\\
&& \Gamma\left( h^J_L - i \frac{\omega^J_L}{2 \pi T^J_L} \right) \Gamma\left( h^J_L + i \frac{\omega^J_L}{2 \pi T^J_L} \right) \Gamma\left( h^J_R - i \frac{\omega^J_R}{2 \pi T^J_R} \right) \Gamma\left( h^J_R + i \frac{\omega^J_R}{2 \pi T^J_R} \right).
\end{eqnarray}
Moreover, since the conformal weights $h^J_L = h^J_R = l + 1$ are integers so
\begin{equation}
\sin\left(\pi h^J_L + i \frac{\omega^J_L}{2 T^J_L} \right) \sin\left(\pi h^J_R + i \frac{\omega^J_R}{2 T^J_R} \right) = (-)^{h^J_L + h^J_R} \sin\left(i \frac{\omega^J_L}{2 T^J_L} \right) \sin\left(i \frac{\omega^J_R}{2 T^J_R} \right).
\end{equation}

From the CFT side, the Euclidean correlator, in terms of the Euclidean frequencies $\omega_{EL} = i \omega_L$, and $\omega_{ER} = i \omega_R$, is
\begin{eqnarray}\label{CFTGR}
G_E(\omega_{EL}, \omega_{ER}) &\sim& T_L^{2h_L -1} T_R^{2h_R - 1} \mathrm{e}^{i \frac{\tilde\omega_{EL}}{2 T_L}} \mathrm{e}^{i \frac{\tilde\omega_{ER}}{2 T_R}}
\nonumber\\
&& \Gamma\left( h_L \!-\! \frac{\tilde\omega_{EL}}{2 \pi T_L} \right) \Gamma\left( h_L \!+\! \frac{\tilde\omega_{EL}}{2 \pi T_L} \right) \Gamma\left( h_R \!-\! \frac{\tilde\omega_{ER}}{2 \pi T_R} \right) \Gamma\left( h_R \!+\! \frac{\tilde\omega_{ER}}{2 \pi T_R} \right),
\end{eqnarray}
where
\begin{equation}
\tilde\omega_{EL} = \omega_{EL} - i q_L \mu_L, \qquad \tilde\omega_{ER} = \omega_{ER} - i q_R \mu_R.
\end{equation}
The retarded Green function $G_R(\omega_L, \omega_R)$ is analytic on the upper half complex $\omega_{L,R}$-plane
\begin{equation}\label{CFTGER}
G_E(\omega_{EL}, \omega_{ER}) = G_R(i \omega_L, i \omega_R), \qquad \omega_{EL}, \omega_{ER} > 0,
\end{equation}
and the Euclidean frequencies $\omega_{EL}$ and $\omega_{ER}$ should take discrete values of the Matsubara frequencies ($m_L, m_R$ are integers for bosons and half integers for fermions)
\begin{equation}\label{Matsubara}
\omega_{EL} = 2 \pi m_L T_L, \qquad \omega_{ER} = 2 \pi m_R T_R.
\end{equation}
At these frequencies, the retarded Green function precisely agrees with the gravity side computation~(\ref{GreenRJ}) up to a numerical normalization factor.

\section{Charge ($Q$-) Picture}

\subsection{Hidden Conformal Symmetry}
For the $Q$-picture CFT$_2$ description one should turn off the $\phi$-direction momentum mode of the charged probe scalar field, by imposing $m = 0$, and the the corresponding radial equation, from Eq.(\ref{eqR}), becomes to
\begin{equation}
\left( \partial_r (\Delta \partial_r) - \frac{\left[ (r_+^2 + a^2) \partial_t + (Q r_+/\ell) \partial_\chi \right]^2}{(r - r_+) (r_+ - r_-)} + \frac{\left[ (r_-^2 + a^2) \partial_t + (Q r_-/\ell) \partial_\chi \right]^2}{(r - r_-) (r_+ - r_-)} \right) \Phi = l (l + 1) \Phi,
\end{equation}
where the operator $\partial_\chi$ acts on ``internal space'' of
$U(1)$ symmetry of the complex scalar field and its eigenvalue is
the charge of the scalar field such as $\partial_\chi \Phi = i \ell
q \Phi$. The parameter $\ell$ in principle can be any values
reflecting an ``ambiguity'' in the holographic dual description of
the RN black hole. Actually, this parameter has a natural
geometrical interpretation as the radius of extra circle when the RN
solution is considered to be embedded into 5D, for more detailed
explanation of $\ell$, see~\cite{Chen:2010bs, Chen:2010yu}. Similarly,
the radial equation is just the Casimir operator~(\ref{Casimir})
acting on $\Phi$ with the following identifications
\begin{eqnarray}
T^Q_L = \frac{(r_+^2 + r_-^2 + 2 a^2) \ell}{4 \pi Q (r_+ r_- - a^2)}, &\quad& T^Q_R = \frac{(r_+^2 - r_-^2) \ell}{4 \pi Q (r_+ r_- - a^2)},
\\
n^Q_L = - \frac{r_+ + r_-}{4 (r_+ r_- - a^2)}, &\quad& n^Q_R = - \frac{r_+ - r_-}{4 (r_+ r_- - a^2)}.
\end{eqnarray}
In this picture, the momentum mode on the $\phi$ direction is turned off and such probe scalar field is not able to explore the information related to the angular momentum. Therefore, it can reveal only the charge dominated subsection resembling the dual CFT of the RN black hole. The corresponding central charges are
\begin{equation}
c^Q_L = c^Q_R = \frac{6 Q^3}{\ell},
\end{equation}
and the CFT microscopic entropy matches the Bekenstein-Hawking area
entropy of the KN black hole (note that $r_+ r_- - a^2 = Q^2$)
\begin{equation}
S^Q_\mathrm{CFT} = \frac{\pi^2}3 \left( c^Q_L T^Q_L + c^Q_R T^Q_R \right) = \pi (r_+^2 + a^2) = S_\mathrm{BH}.
\end{equation}

\subsection{Scattering}
As what we did for the $J$-picture, we are going to study the scattering process to find further supports for the duality. In terms of new coordinate $z$ defined in~(\ref{defZ}), the solutions of the near region KG equation~(\ref{eqR}), with $m = 0$, also includes ingoing and outgoing modes
\begin{eqnarray}
R_Q^\mathrm{(in)} &=& z^{- i \gamma_Q}  (1 - z)^{l + 1} \, F( a_Q, b_Q; c_Q; z ),
\nonumber\\
R_Q^\mathrm{(out)} &=& z^{i \gamma_Q}  (1 - z)^{l + 1} \, F( a_Q^*, b_Q^*; c_Q^*; z ),
\end{eqnarray}
where the coefficients are
\begin{eqnarray}
\gamma_Q &=& \frac{\omega (r_+^2 + a^2) - q Q r_+}{r_+ - r_-},
\nonumber\\
a_Q &=& 1 + l - i \frac{\omega (r_+^2 + r_-^2 + 2 a^2) - q Q (r_+ + r_-)}{r_+ - r_-},
\nonumber\\
b_Q &=& 1 + l - i \left[ \omega (r_+ + r_-) - q Q \right],
\nonumber\\
c_Q &=& 1 - i 2 \gamma_Q.
\end{eqnarray}
Again, the relation $c_Q - a_Q - b_Q = - 2 l - 1$ holds. The crucial asymptotically behavior of the ingoing mode at the match region $r \gg M$ (but still satisfying $r \ll 1/\omega$) can be obtained by taking the limits $z \to 1$ and $1 - z \to r^{-1}$, and the result is
\begin{equation} \label{asymRQ}
R_Q^\mathrm{(in)}(r \gg M) \sim A_Q \, r^l + B_Q \, r^{- l - 1},
\end{equation}
where the two key coefficients are
\begin{equation}
A_Q = \frac{\Gamma(c_Q) \Gamma(2 l + 1)}{\Gamma(a_Q) \Gamma(b_Q)}, \qquad B_Q = \frac{\Gamma(c_Q) \Gamma(- 2 l - 1)}{\Gamma(c_Q - a_Q) \Gamma(c_Q - b_Q)}.
\end{equation}
As like the case in the $J$-picture, the conformal weights of the operator dual to the scalar field are
\begin{equation}
h^Q_L = h^Q_R = l + 1.
\end{equation}
Hence, the coefficients $a_Q$ and $b_Q$ can be expressed in terms of conformal weights (real part) and two parameters $\tilde\omega^Q_L$ and $\tilde\omega^Q_R$ (imaginary part)
\begin{equation}
a_Q = h^Q_R - i \frac{\tilde\omega^Q_R}{2 \pi T^Q_R}, \qquad b_Q = h^Q_L - i \frac{\tilde\omega^Q_L}{2 \pi T^Q_L},
\end{equation}
where, unlike the $J$-picture, $\tilde\omega^Q_L, \tilde\omega^Q_R$
are composed by three sets CFT parameters: frequencies ($\omega^Q_L,
\omega^Q_R$), charges ($q^Q_L, q^Q_R$) and chemical potentials
($\mu^Q_L, \mu^Q_R$) as
\begin{equation}
\tilde\omega^Q_L = \omega^Q_L - q^Q_L \mu^Q_L, \qquad \tilde\omega^Q_R = \omega^Q_R - q^Q_R \mu^Q_R,
\end{equation}
with
\begin{eqnarray}
&& \omega^Q_L = \frac{\ell \omega (r_+ + r_-)(r_+^2 + r_-^2 + 2a^2)}{2 Q (r_+ r_- - a^2)}, \qquad q^Q_L = q, \qquad \mu^Q_L = \frac{\ell (r_+^2 + r_-^2 + 2a^2)}{2 (r_+ r_- - a^2)},
\nonumber\\
&& \omega^Q_R = \frac{\ell \omega (r_+ + r_-)(r_+^2 + r_-^2 + 2a^2)}{2 Q (r_+ r_- - a^2)}, \qquad q^Q_R = q, \qquad \mu^Q_R = \frac{\ell (r_+ + r_-)^2}{2 (r_+ r_- - a^2)}.
\end{eqnarray}
The essential part of the absorption cross section can be read out directly from the coefficient $A_Q$, namely
\begin{equation}
P^Q_\mathrm{abs} \sim |A_Q|^{-2} \sim \sinh( 2 \pi \gamma_Q) ) \, \left| \Gamma(a_Q) \right|^2 \, \left| \Gamma(b_Q) \right|^2.
\end{equation}
Again, the conjugate charges in this picture
\begin{equation}
\delta S^Q_{CFT} = \frac{\delta E^Q_L}{T^Q_L} + \frac{\delta E^Q_R}{T^Q_R},
\end{equation}
can be solved according to the first law of black hole thermodynamics~(\ref{1st}) and relation $\delta S^Q_{CFT} = \delta S_{BH}$. The solution is
\begin{eqnarray}
\delta E^Q_L &=& \frac{2 \ell (2M^2 - Q^2) M}{Q^3} \delta M - \ell \left( \frac{2 M^2}{Q^2} - 1 \right) \delta Q,
\nonumber\\
\delta E^Q_R &=& \frac{2 \ell (2M^2 - Q^2) M}{Q^3} \delta M - \frac{2 \ell M^2}{Q^2} \delta Q - \frac{2 \ell J}{Q^3} \delta J,
\end{eqnarray}
The identifications of parameters are $\delta M = \omega, \delta Q = q$ and $\delta J = m$. Since the probe scalar does not have momentum mode along $\phi$-direction, in such case, we have
\begin{equation}
\tilde\omega^Q_L = \delta E^Q_L(\delta M = \omega, \delta Q = q; \delta J = 0), \qquad \tilde\omega^Q_R = \delta E^Q_R(\delta M = \omega, \delta Q = q; \delta J = 0).
\end{equation}

Therefore, one can straightforwardly verify the following relation for the imaginary part of the coefficient $c_Q$,
\begin{equation}
2 \pi \gamma_Q = \frac{\tilde\omega^Q_L}{2 T^Q_L} + \frac{\tilde\omega^Q_R}{2 T^Q_R}.
\end{equation}
Finally, the absorption cross section can be expressed as
\begin{equation}
P^Q_\mathrm{abs} \sim (T^Q_L)^{2 h^Q_L - 1} (T^Q_R)^{2 h^Q_R - 1} \sinh\left( \frac{\tilde\omega^Q_L}{2 T^Q_L} + \frac{\tilde\omega^Q_R}{2 T^Q_R} \right) \left| \Gamma\left( h^Q_L + i \frac{\tilde\omega^Q_L}{2 \pi T^Q_L} \right) \right|^2 \, \left| \Gamma\left( h^Q_R + i \frac{\tilde\omega^Q_R}{2 \pi T^Q_R} \right) \right|^2,
\end{equation}
which has the same form of the finite temperature absorption cross
section of an operator with the conformal weights ($h_L, h_R$), frequencies ($\omega_L, \omega_R$)
electric charges ($q_L, q_R$) and chemical potentials ($\mu_L, \mu_R$)
in the dual 2D CFT with the temperatures ($T_L, T_R$).

\subsection{Real-time Correlator}
In the $Q$-picture, the two-point retarded correlator is simple as well
\begin{equation}
G^Q_R \sim \frac{B_Q}{A_Q} = \frac{\Gamma(- 2 l - 1)}{\Gamma(2 l + 1)} \frac{\Gamma(a_Q) \Gamma(b_Q)}{\Gamma(c_Q - a_Q) \Gamma(c_Q - b_Q)},
\end{equation}
Repeating the same analysis done in the $J$-picture, we firstly can easily check that retarded Green function, by following the identity $c_Q - a_Q - b_Q = - 2 l - 1$, is
\begin{equation}
G^Q_R \sim
\frac{\Gamma\left( h^Q_L - i \frac{\tilde\omega^Q_L}{2 \pi T^Q_L} \right) \Gamma\left( h^Q_R - i \frac{\tilde\omega^Q_R}{2 \pi T^Q_R} \right)}{\Gamma\left( 1 - h^Q_L - i \frac{\tilde\omega^Q_L}{2 \pi T^Q_L} \right) \Gamma\left( 1 - h^Q_R - i \frac{\tilde\omega^Q_R}{2 \pi T^Q_R} \right)}.
\end{equation}
Then, by using the relation $\Gamma(z) \Gamma(1-z) = \pi/\sin(\pi z)$ we have
\begin{eqnarray}\label{GreenRQ}
G^Q_R &\sim& \sin\left(\pi h^Q_L + i \frac{\tilde\omega^Q_L}{2 T^Q_L} \right) \sin\left(\pi h^Q_R + i \frac{\tilde\omega^Q_R}{2 T^Q_R} \right)
\nonumber\\
&& \Gamma\left( h^Q_L - i \frac{\tilde\omega^Q_L}{2 \pi T^Q_L} \right) \Gamma\left( h^Q_L + i \frac{\tilde\omega^Q_L}{2 \pi T^Q_L} \right) \Gamma\left( h^Q_R - i \frac{\tilde\omega^Q_R}{2 \pi T^Q_R} \right) \Gamma\left( h^Q_R + i \frac{\tilde\omega^Q_R}{2 \pi T^Q_R} \right),
\end{eqnarray}
and, since the conformal weights $h^Q_L = h^Q_R = l + 1$ are integers so that
\begin{equation}
\sin\left(\pi h^Q_L + i \frac{\tilde\omega^Q_L}{2 T^Q_L} \right) \sin\left(\pi h^Q_R + i \frac{\tilde\omega^Q_R}{2 T^Q_R} \right) = (-)^{h^Q_L + h^Q_R} \sin\left(i \frac{\tilde\omega^Q_L}{2 T^Q_L} \right) \sin\left(i \frac{\tilde\omega^Q_R}{2 T^Q_R} \right).
\end{equation}
At the Matsubara frequencies~(\ref{Matsubara}), the retarded Green function precisely agrees with the CFT results~(\ref{CFTGR}-\ref{CFTGER}) up to a normalization factor depending on $q^Q_L$ and $q^Q_R$.

\section{Conclusion}
It is a remarkable result that there are two different individual 2D
CFTs holographically dual to the KN black hole. In fact this is an
expectable result since from the gravity side, the KN black hole
will return to the Kerr black hole when $Q = 0$ while to the RN
black hole when $J = 0$. Thus it is probable to describe the quantum
gravity description of the KN black hole in terms of the Kerr/CFT or
RN/CFT$_2$ dualities. In this paper, we have explored the twofold
hidden 2D conformal symmetries in the generic non-extremal KN black
hole by analyzing the near region wave equation of a probe scalar
field at low frequencies. Despite the fact that these 2D conformal
symmetries are not derived directly from the KN black hole geometry,
they should reflect the information of the background. The geometric
picture of the apparent AdS$_3 \sim$ AdS$_2 \times S^1$ structure
could be understood in the follows. The radial and the time
coordinates (associated to the frequency mode $\omega$) of the KN
black hole serve as the AdS$_2$ base manifold. Besides, there are
two possible choices for the $U(1)$ fiber to form an apparent
AdS$_3$ structure: either the $\phi$-coordinate (associated with the
momentum mode $m$) or the internal phase of the probe scalar field
(associated with the electric charge $q$), as shown in Figure 1
(This geometric picture becomes more faithful in the near horizon
near extremal limit of the KN black hole). It is shown that each
choice gives an individual CFT$_2$ description, called the
$J$-picture for the first choice coupled to the angular momentum,
and the $Q$-picture for the second one coupled to the background
gauge field, dual to the KN black hole. In both pictures, the CFT
central charges are transcribed exactly from their ``descendants'',
i.e. the Kerr/CFT and the RN/CFT$_2$ correspondences, respectively.
These twofold dualities are further supported by the agreements of
the entropies, absorption cross sections and real time correlators
computed from both the gravity and the CFT sides.

\begin{figure}[ht]\label{fig1}
\includegraphics[width=10cm]{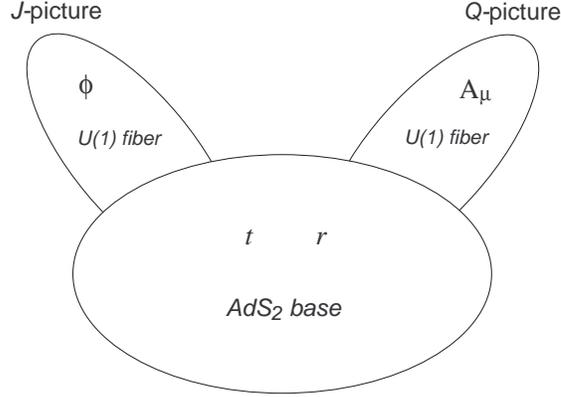}
\caption{Twofold conformal symmetries associated with the Kerr-Newman black holes.}
\end{figure}

Therefore, in addition to the mass, every one of the other two
macroscopic hairs of the KN black hole, either the angular momentum
or the charge, can provide an individual holographically dual
CFT$_2$ description.
Actually, it has been discussed that for the higher dimensional
rotating black holes, each angular momentum can provide an
individual dual CFT$_2$, see for example~\cite{Lu:2008jk,
Chen:2009xja}. According to our results, the charge of black holes
also shares the same property. Hence, we conjecture that there
should be a general property for any black hole sourced by abelian
gauge fields, which could be summarized as a ``microscopic no hair
theorem'': {\em for each macroscopic hair parameter, in additional
to the mass of a black hole in the Einstein-Maxwell theory (with
dimension $D \geq 3$), there should exist an associated holographic
CFT$_2$ description.}

\section*{Acknowledgement}
This work was supported by the National Science Council of the R.O.C. under the grant NSC 96-2112-M-008-006-MY3 and in part by the National Center of Theoretical Sciences (NCTS).

\begin{appendix}
\section{Symmetry and Casimir Operator of $\mathrm{AdS}_3$}
The two sets of symmetry generators of the AdS$_3$ space with radius $L$, in the Poincar\'e coordinates: ($w^\pm, y$),
\begin{equation}\label{dsAdS3}
ds_3^2 = \frac{L^2}{y^2} ( dy^2 + dw^+ dw^-),
\end{equation}
are
\begin{eqnarray}
H_1 = i \partial_+, \quad H_0 = i \left( w^+ \, \partial_+ + \frac12 y \, \partial_y \right), \quad H_{-1} = i \left( (w^+)^2 \, \partial_+ + w^+ y \, \partial_y - y^2 \, \partial_- \right),
\\
\bar H_1 = i \partial_-, \quad \bar H_0 = i \left( w^- \, \partial_- + \frac12 y \, \partial_y \right), \quad \bar H_{-1} = i \left( (w^-)^2 \, \partial_- + w^- y \, \partial_y - y^2 \, \partial_+ \right),
\end{eqnarray}
assembling two copies of the $SL(2,R)$ Lie algebra
\begin{equation}
\left[ H_0, H_{\pm 1} \right] = \mp i H_{\pm 1}, \qquad \left[ H_{-1}, H_1 \right] = - 2 i H_0.
\end{equation}
Thus the corresponding Casimir operator is
\begin{equation}
\mathcal{H}^2 = \bar\mathcal{H}^2 = - H_0^2 + \frac12 \left( H_1 H_{-1} + H_{-1} H_1 \right) = \frac14 \left( y^2 \, \partial_y^2 - y \, \partial_y \right) + y^2 \, \partial_+ \partial_-.
\end{equation}

Converting the Poincar\'e coordinates ($w^\pm, y$) to the coordinates ($t, r, \chi$) by the following transformations
\begin{eqnarray}\label{wy2tr}
w^+ &=& \sqrt{\frac{r - r_+}{r - r_-}} \, \exp(2 \pi T_R \chi + 2 n_R t),
\nonumber\\
w^- &=& \sqrt{\frac{r - r_+}{r - r_-}} \, \exp(2 \pi T_L \chi + 2 n_L t),
\nonumber\\
y &=& \sqrt{\frac{r_+ - r_-}{r - r_-}} \, \exp[\pi (T_R + T_L) \chi + (n_R + n_L) t],
\end{eqnarray}
we can directly calculate all the $SL(2,R)$ generators in terms of black hole coordinates
\begin{eqnarray*}
H_1 &=& i \mathrm{e}^{-(2 \pi T_R \chi + 2 n_R t)} \left[ \sqrt{\Delta} \partial_r \!+\! \frac{n_L (\delta_- \!+\! \delta_+) \!+\! n_R (\delta_- \!-\! \delta_+)}{4 \pi \sqrt{\Delta} \mathcal{A}} \partial_\chi \!-\! \frac{T_L (\delta_- \!+\! \delta_+) \!+\! T_R (\delta_- \!-\! \delta_+)}{4 \sqrt{\Delta} \mathcal{A}} \, \partial_t \right],
\nonumber\\
H_0 &=& i \left[ \frac{n_L}{2 \pi \mathcal{A}} \, \partial_\chi - \frac{T_L}{2 \mathcal{A}} \, \partial_t \right],
\nonumber\\
H_{-1} &=& i \mathrm{e}^{2 \pi T_R \chi + 2 n_R t} \left[ -\sqrt{\Delta} \partial_r \!+\! \frac{n_L (\delta_- \!+\! \delta_+) \!+\! n_R (\delta_- \!-\! \delta_+)}{4 \pi \sqrt{\Delta} \mathcal{A}} \partial_\chi \!-\! \frac{T_L (\delta_- \!+\! \delta_+) \!+\! T_R (\delta_- \!-\! \delta_+)}{4 \sqrt{\Delta} \mathcal{A}} \, \partial_t \right],
\end{eqnarray*}
and
\begin{eqnarray*}
\bar H_1 &=& i \mathrm{e}^{-(2 \pi T_L \chi + 2 n_L t)} \left[ \sqrt{\Delta} \partial_r \!-\! \frac{n_R (\delta_- \!+\! \delta_+) \!+\! n_L (\delta_- \!-\! \delta_+)}{4 \pi \sqrt{\Delta} \mathcal{A}} \partial_\chi \!+\! \frac{T_R (\delta_- \!+\! \delta_+) \!+\! T_L (\delta_- \!-\! \delta_+)}{4 \sqrt{\Delta} \mathcal{A}} \, \partial_t \right],
\nonumber\\
\bar H_0 &=& i \left[ - \frac{n_R}{2 \pi \mathcal{A}} \, \partial_\chi - \frac{T_R}{2 \mathcal{A}} \, \partial_t \right],
\nonumber\\
\bar H_{-1} &=& i \mathrm{e}^{2 \pi T_L \chi + 2 n_L t} \left[ - \sqrt{\Delta} \partial_r \!-\! \frac{n_R (\delta_- \!+\! \delta_+) \!+\! n_L (\delta_- \!-\! \delta_+)}{4 \pi \sqrt{\Delta} \mathcal{A}} \partial_\chi \!+\! \frac{T_R (\delta_- \!+\! \delta_+) \!+\! T_L (\delta_- \!-\! \delta_+)}{4 \sqrt{\Delta} \mathcal{A}} \, \partial_t \right],
\end{eqnarray*}
where
\begin{equation}
\delta_\pm = r - r_\pm, \qquad \mathcal{A} = T_R n_L - T_L n_R.
\end{equation}
Finally the Casimir operator becomes
\begin{equation}\label{Casimir}
\mathcal{H}^2 = \partial_r (\Delta \partial_r) - \frac{r_+ - r_-}{r - r_+} \left( \frac{T_L + T_R}{4 \mathcal{A}} \partial_t - \frac{n_L + n_R}{4 \pi \mathcal{A}} \partial_\chi \right)^2 + \frac{r_+ - r_-}{r - r_-} \left( \frac{T_L - T_R}{4 \mathcal{A}} \partial_t - \frac{n_L - n_R}{4 \pi \mathcal{A}} \partial_\chi \right)^2.
\end{equation}
\end{appendix}


\end{document}